\documentclass[%
 reprint,
 amsmath,amssymb,
 aps,
]{revtex4-2}

\usepackage{arydshln}
\usepackage{bm}

\usepackage{graphicx}
\usepackage{dcolumn}
\usepackage{bm}
\usepackage{txfonts}
\usepackage{caption}
\usepackage{subcaption}
\usepackage{cleveref}

\begin{document}

\title{The ABJM Momentum Amplituhedron \\ 
	\normalsize ABJM Scattering Amplitudes From Configurations of Points in Minkowski Space}

\author{Tomasz {\L}ukowski}
\email{t.lukowski@herts.ac.uk}
 \affiliation{Department of Physics, Astronomy and Mathematics, University of Hertfordshire,\\AL10 9AB Hatfield, Hertfordshire, United Kingdom}
\author{Jonah Stalknecht}%
 \email{j.stalknecht@herts.ac.uk}
\affiliation{%
 Department of Physics, Astronomy and Mathematics, University of Hertfordshire,\\AL10 9AB Hatfield, Hertfordshire, United Kingdom
}%

\date{\today}

\begin{abstract}
In this paper, we define the ABJM loop momentum amplituhedron, which is a geometry encoding ABJM planar tree-level amplitudes and loop integrands in the three-dimensional spinor helicity space. Translating it to the space of dual momenta produces a remarkably simple geometry given by configurations of space-like separated off-shell momenta living inside a curvy polytope defined by momenta of scattered particles. We conjecture that the canonical differential form on this space gives amplitude integrands, and we provide a new formula for all one-loop $n$-particle integrands in the positive branch. For higher loop orders, we utilize the causal structure of configurations of points in Minkowski space to explain the singularity structure for known results at two loops.
\end{abstract}

\maketitle

\section{Introduction}
Recent years have seen remarkable progress in applying positive geometries \cite{Arkani-Hamed:2017tmz} to the problem of finding scattering amplitudes in $\mathcal{N}=4 $ super Yang-Mills (sYM) \cite{Arkani-Hamed:2013jha,Arkani-Hamed:2013kca,Arkani-Hamed:2017vfh,Franco:2014csa,Herrmann:2020qlt,Ferro:2015grk,Ferro:2018vpf,Damgaard:2019ztj,Ferro:2022abq,Damgaard:2021qbi,Damgaard:2020eox,Ferro:2020lgp,Lukowski:2022fwz,Lukowski:2019kqi,Lukowski:2020dpn,Lukowski:2021amu, Parisi:2021oql,Karp:2016uax,Karp:2017ouj,Arkani-Hamed:2018rsk,YelleshpurSrikant:2019meu,Langer:2019iuo,Bai:2015qoa,Kojima:2018qzz,Dian:2022tpf} and ABJM \cite{He:2021llb,Huang:2021jlh,He:2023rou,Lukowski:2021fkf,He:2022cup,He:2023exb,Henn:2023pkc} theories, based on previous works on positive Grassmannians \cite{Postnikov:2006kva,Arkani-Hamed:2009ljj,Arkani-Hamed:2012zlh,Lee:2010du,Elvang:2014fja,Huang:2013owa,Huang:2014xza,Kim:2014hva}. Most recently, by considering the reduction of the kinematics from the four-dimensional space of massless momenta to three dimensions, the ABJM amplituhedron $\mathcal{W}_n^{(L)}$ was defined in \cite{He:2023rou} in the three-dimensional momentum twistor space. This geometry encodes planar tree-level amplitudes $A_{n}^{(0)}$ and loop integrands $A_n^{(L)}$ in its canonical differential form. Importantly, it was conjectured in \cite{He:2023rou} that the $L$-loop integrands can be explicitly obtained from the ABJM amplituhedron by subdividing it into smaller pieces that are cartesian products of tree-level geometry ($\mathcal{C}_m$) times the $L$-loop geometry ($\mathcal{L}^{(L)}_m$)
\begin{equation}\label{eq:factorisation_chambers}
\mathcal{W}_n^{(L)}=\bigcup_m \mathcal{C}_m \times \mathcal{L}^{(L)}_m\,,
\end{equation}
where $\mathcal{C}_m$ are maximal intersections of BCFW cells at tree level termed {\it chambers}. For a given chamber, the loop geometry is the same, and can be though of as a fibration of the loop amplituhedron over the tree one.

In this paper we define a close cousin to the ABJM amplituhedron we termed the {\it ABJM momentum amplituhedron} $\mathcal{A}_n^{(L)}$, that lives directly in the three-dimensional spinor helicity space. To do that, we utilize the map used in the recently conjectured construction of the loop momentum amplituhedron for $\mathcal{N}=4$ sYM \cite{Ferro:2022abq}. This will define a geometry whose differential form is the integrand for the so-called {\it positive branch} of the theory. Importantly, the geometry that we obtain is remarkably basic when depicted in the space of dual momenta in three-dimensional Minkowski space. In particular, for a given tree-level configuration of points in $\mathcal{C}_m$, the loop geometry is a subset of the Cartesian product of $L$ curvy versions of simple polytopes we denote $\Delta_{n}^{(m)}$. Remarkably, at one-loop the ABJM loop momentum amplituhedron in a given chamber {\it is} the curvy polytope $\Delta_{n}^{(m)}$, and it is straightforward to find its canonical form. For each chamber one can easily identify vertices of $\Delta_{n}^{(m)}$ and therefore find a general form of all one-loop integrands $A_{n}^{(1)}$ in the positive branch for scattering amplitudes with $n=2k$ particles:
\begin{align}\label{eq:oneloopall}
A_{n}^{(1)}&=\sum_{(T_1,T_2)\in \mathcal{T}^{o}\times \mathcal{T}^{e}}\Omega_{T_1,T_2}^{(0)}\wedge\Omega^{(1)}_{T_1,T_2}\,.
\end{align}
Here $\Omega_{T_1,T_2}^{(0)}$ is the tree-level canonical form for a given chamber that we labelled by a pair of triangulations $(T_1,T_2)$ of two $k$-gons formed of odd/even particle labels. The one-loop differential form $\Omega^{(1)}_{T_1,T_2}$ associated with the chamber $(T_1,T_2)$ takes the form
\begin{align*}
\Omega^{(1)}_{T_1,T_2}&=\sum_{a=1}^n (-1)^a \omega_{a-1,a,a+1}+\sum_{(a,b,c)\in T_1}\omega^+_{abc}-\sum_{(a,b,c)\in T_2}\omega^-_{abc}\,,
\end{align*}
and we present its explicit expression later. Results for other branches can be obtained from \eqref{eq:oneloopall} by parity operations defined in \cite{He:2023rou}.

For higher loops, the ABJM momentum amplituhedron $\mathcal{A}_n^{(L)}$ is specified by configurations of $L$ points inside $\Delta_{n}^{(m)}$ that are space-like separated from each other. By studying such configurations of points, we are able to give a simple explanation for the structure of the answer for the two-loop integrands known for $n=4$ \cite{Chen:2011vv, Bianchi:2011dg}, $n=6$ \cite{Caron-Huot:2012sos}\ and $n=8$ \cite{He:2022lfz}. To do that, we will utilise the notion of negative geometries \cite{Arkani-Hamed:2021iya} and use the causal structure of the three-dimensional Minkowski space. 

This paper is organised as follows: we start by recalling basic facts about three-dimensional Minkowski space that will set the stage for next sections. Then, we study configurations of dual momenta that originate from the definition of the tree-level ABJM momentum amplituhedron that will allow us to define the curvy polytopes $\Delta_{n}^{(m)}$. We follow by defining the ABJM momentum amplituhedron at loop level, and detailing its structure at one loop. In particular, we provide the explicit formula for one-loop integrands in the positive branch for all multiplicities. The final section focuses on the two-loop geometry. We conclude the paper with some open questions arising from our construction.

\section{ABJM momentum amplituhedron\label{sec:level1}}
\subsection{Three-dimensional Minkowski Space}
We will work in the three-dimensional Minkowski space $\mathcal{M}$ with signature $(+,-,-)$. Scattering data for $n$-particle scattering in ABJM is encoded in a set of $n=2k$ three-dimensional on-shell momenta $p_a^\mu$, $a=1,\ldots,2k$, $\mu=0,1,2$, with $(p_a)^2=0$. We assume that particles with odd labels are outgoing and the ones with even labels are incoming. This leads to the following momentum conservation
\begin{equation}
\sum_{a\, \text{odd}}p_a^\mu-\sum_{a\, \text{even}}p_a^\mu=0\,.
\end{equation} 
In planar theory, this data can be equivalently encoded using dual coordinates
\begin{equation}\label{eq.dual}
p_a^\mu \eqqcolon x_{a+1}^\mu-x_{a}^\mu\,,
\end{equation}
that define a null polygon in Minkowski space. For convenience, we choose $x_1=0$. This allows us to invert relation \eqref{eq.dual} to get
\begin{equation}
x_b=\sum_{a=1}^{b-1}(-1)^ap_a\,.
\end{equation}
We denote by $\mathcal{I}_x=\{y\in\mathcal{M}:(x-y)^2=0\}$ the lightcone of point $x$. 

The on-shell condition $p^2=0$ can be resolved by introducing three-dimensional spinor helicity variables and writing
\begin{equation}
p^{\alpha\beta}=\begin{pmatrix}
-p^0+p^2&p^1\\p^1&-p^0-p^2
\end{pmatrix}\eqqcolon\lambda^\alpha\lambda^\beta\,.
\end{equation}
Scattering amplitudes are invariant under the action of the Lorentz group $SL(2)$ on $\lambda$ and therefore a point in the kinematic space is an element of the orthogonal Grassmannian $\lambda\in OG(2,2k)\eqqcolon\mathcal{K}_{2k}$, where orthogonality is defined with respect to $\eta=\mathrm{diag}(+,-,\ldots,+,-)$. In the following, we will repeatedly use the spinor brackets $\langle ab\rangle:=\lambda_a^1\lambda_b^2-\lambda_a^2\lambda_b^1$.

\subsection{Tree-level momentum amplituhedron}

Following \cite{Huang:2021jlh,He:2021llb}, the tree-level ABJM momentum amplituhedron $\mathcal{A}_{2k}^{(0)} =\phi_\Lambda(OG_+(k,2k))$ is a subset of the kinematic space $\mathcal{K}_{2k}$ which is the image of the positive orthogonal Grassmannian $OG_+(k,2k)$ through the map

\begin{align}\label{eq:lambda}
\begin{array}{cccl}
\phi_\Lambda\colon&OG_+(k,2k)&\to& OG(2,2k)\\
&C&\mapsto &\lambda=(C\cdot \eta\cdot\Lambda)^\perp \cdot\Lambda\,,
\end{array}
\end{align}
where $\Lambda\in M(k+2,2k)$ is a fixed matrix  \footnote{We use a modified version of the definition in \cite{Huang:2021jlh,He:2021llb} that can be mapped to the original one by the parity duality.}. To get the desired geometry, in this paper we slightly modify the original definition and demand that the matrix $\Lambda$ is such that the image contains points satisfying the following sign-flip pattern: $\langle ii+1\rangle>0$ for all $i=1,\ldots,n$, the sequence $\{\langle 12\rangle,\langle 13\rangle,\ldots,\langle 1n\rangle\}$ has $k-2$ sign flips and all planar Mandelstams are negative. We found that taking a generic twisted positive matrix $\Lambda$, i.e. $\Lambda \cdot \eta$ has all maximal minors positive, leads to the correct sign-flip pattern. 

Importantly, for every point $\lambda\in \mathcal{A}_{2k}^{(0)}$, when translated to the dual space we get a configuration of dual points $x_a$, $a=1,\ldots,n$ forming a null polygonal loop, that  satisfy the conditions
\begin{equation}
\,\quad (x_a-x_b)^2<0,\, \mathrm{for}\, |a-b|>1\,,
\end{equation}
and all even-indexed points are in the future of their (null-separated)  odd-indexed two neighbours.

\begin{figure}
\begin{center}
\includegraphics[height=32mm]{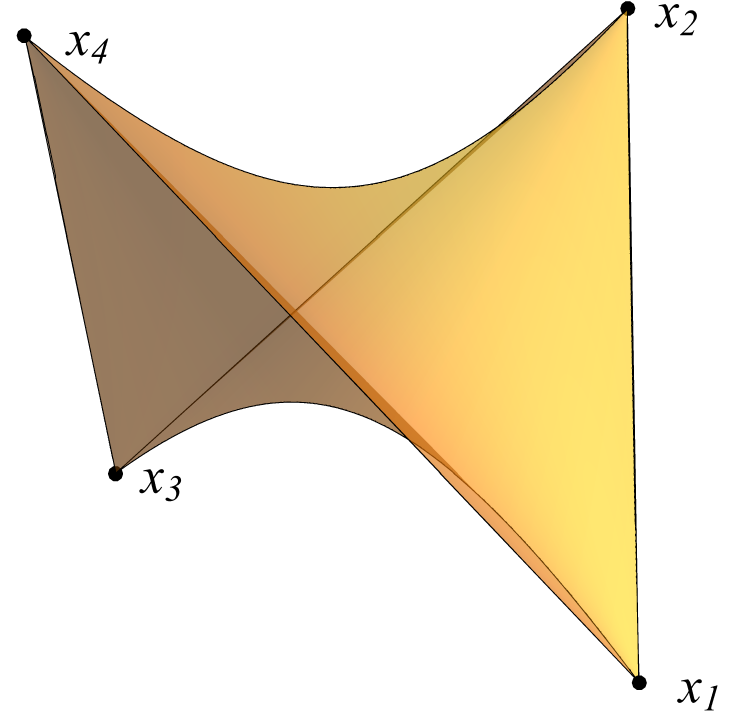}\quad
\includegraphics[height=32mm]{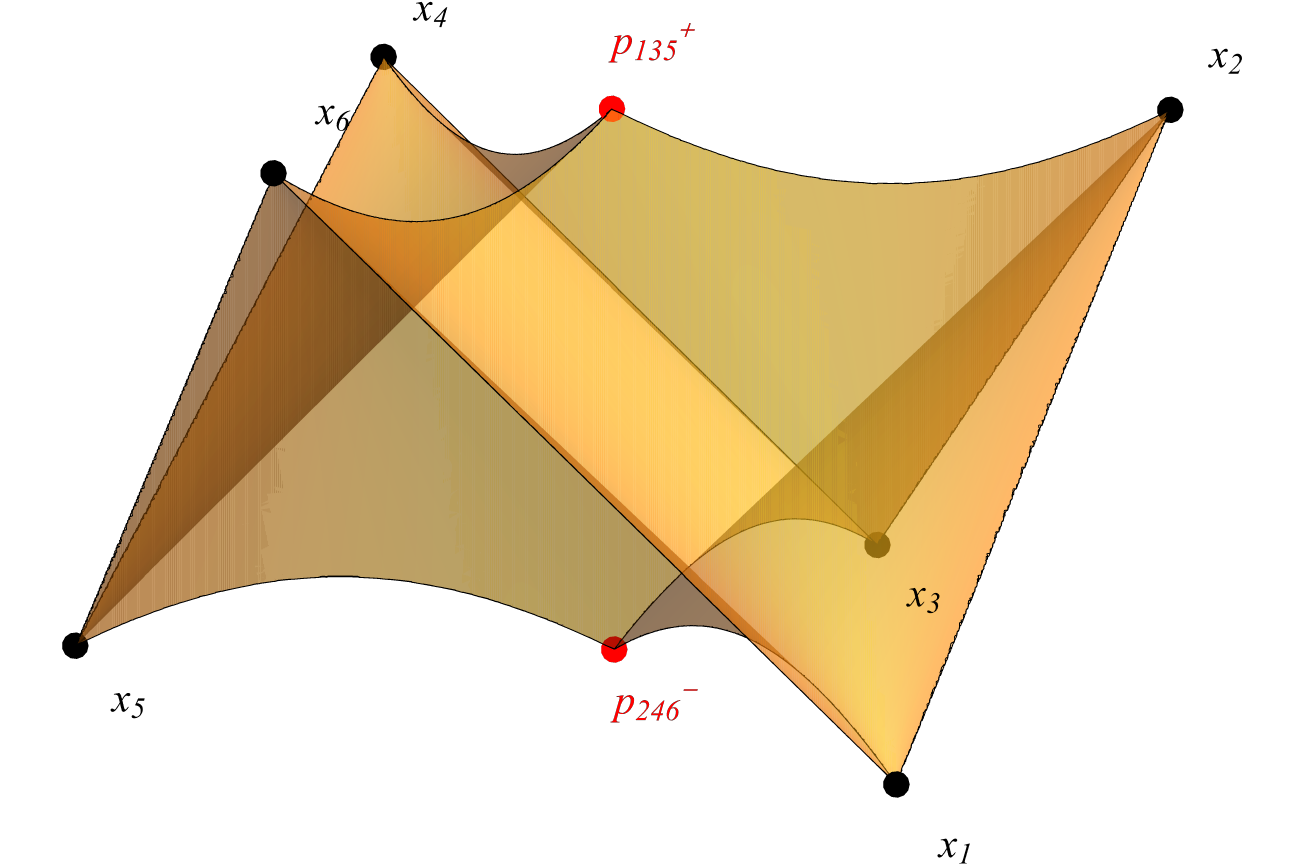}
\end{center}
\caption{The region $\Delta_n$ that is the compact part of the set of all points that are space-like separated from $x_a$ for $n=4$ and $n=6$.}
\label{fig:n46}
\end{figure}

These configurations of points $x_a$ have a very interesting and intricate structure. In particular, for any three generic points $x_a$, $x_b$ and $x_c$ that are not neighbours, we find two points in the intersection of their lightcones $\mathcal{I}_{x_a}\cap \mathcal{I}_{x_b}\cap\mathcal{I}_{x_c}$, one in the future of points $(x_a,x_b,x_c)$, and one in the past. We denote the future (past) point as $p_{abc}^+$ ($p_{abc}^-$).
Motivated by our future considerations, let us define the region of the Minkowski space
\begin{equation}
\mathcal{K}^{\leq0}_{x_a}\coloneqq\{ x\in \mathcal{M}\colon(x-x_a)^2\leq0\,\,\mathrm{for}\,\,a=1,\ldots ,n\}\,,
\end{equation}
containing all points that
 are space-like (or light-like) separated from all $x_a$. This region is non-empty since $x_a\in \mathcal{K}_{x_a}^{\leq0}$ for all $a=1,\ldots,n$. Moreover, it can be naturally divided into two pieces: a compact one that we denote $\Delta_{n}(x_a)$ and a non-compact one $\overline{\Delta}_{n}(x_a)$. We depict the compact region for $n=4$ and $n=6$ in figure \ref{fig:n46}.
It is easy to see that the region $\Delta_{4}(x_a)$ is a curvy version of a tetrahedron, while $\Delta_{6}(x_a)$ is a curvy version of a cube. The latter has two vertices coming from triple intersections of lighcones, $p_{135}^+$ and $p_{246}^-$, in addition to the six vertices $x_i$. For higher $n$, the shape of $\Delta_{n}(x_a)$ is more involved and starts to depend on the details of the configuration of $x$'s. For example, for $n=8$ there are four distinct geometries $\Delta_{8}^{(m)}(x_a)$, each of them contains eight vertices $x_a$ together with triple intersections of lightcones:
\begin{align*}
\mathcal{V}(\Delta_{8}^{(1)}(x_a))=& \{x_a,p_{135}^+,p_{157}^+,p_{246}^-,p_{268}^-\}\,,\\
\mathcal{V}(\Delta_{8}^{(2)}(x_a))=& \{x_a,p_{137}^+,p_{357}^+,p_{246}^-,p_{268}^-\}\,,\\
\mathcal{V}(\Delta_{8}^{(3)}(x_a))=& \{x_a,p_{135}^+,p_{157}^+,p_{248}^-,p_{468}^-\}\,,\\
\mathcal{V}(\Delta_{8}^{(4)}(x_a))=& \{x_a,p_{137}^+,p_{357}^+,p_{248}^-,p_{468}^-\}\,.
\end{align*}
These four cases exactly correspond to the four chambers for $n=8$ in \cite{He:2023rou}! One notices that the labels of the intersecting cones for each chamber can be thought of as products of triangulations of two 4-gons: one formed of odd labels, and one formed of even labels. This pattern continues for higher $n$. We found that there are exactly $C_{k-2}^2$ different geometries $\Delta_{n}^{(m)}$ for $n=2k$ particles, where $C_p$ is the $p$\textsuperscript{th} Catalan number. All these geometries have $n$ vertices corresponding to $x_a$ together with exactly $(k-2)$ triple intersections of lightcones of points with odd labels, and $(k-2)$ intersections with even labels. The geometry changes at non-generic configurations of points $x_a$, where an intersection of four lighcones is possible, corresponding to a bistellar flip in one of the aforementioned triangulations. All curvy polytopes $\Delta_n^{(m)}$ are {\it simple}, meaning that each vertex has exactly three edges originating from it.

We note that the two triangulations of $k$-gons are more than just a mnemonic to distinguish distinct geometries: they in fact provide part of the skeleton of the dual geometry of $\smash{\Delta_{n}^{(m)}}$! The fact that the facets of this dual are all triangles is equivalent to the statement that polytopes $\Delta_{n}^{(m)}$ are simple. The triangles with vertices $(a,b,c)$ in the dual correspond to vertices $p_{abc}$ of $\smash{\Delta_{n}^{(m)}}$. Since the dual of a triangulation of a $k$-gon is a planar tree Feynman diagram for $k$ particles, the skeleton of our geometry is that of a null polygon between points $x_a$, with two distinct Feynman diagrams drawn between the odd and even points, where different choices of Feynman diagrams label different chambers. Therefore, distinct $\Delta_{n}^{(m)}$ correspond to all pairs of two planar tree Feynman diagrams for $k$ particles. 

\subsection{Loop-level momentum amplituhedron}
Given a fixed tree-level configuration $\lambda\in \mathcal{A}_n^{(0)}$, we will define the ABJM loop momentum amplituhedron using the map 
\begin{align}\label{eq:phi_lambda}
\begin{array}{cccc}
\Phi_\lambda\colon&G(2,2k)^L&\to& GL(2)^L\\
&(D_1,\ldots,D_L)&\mapsto &(\ell_1,\ldots,\ell_L)
\end{array}
\end{align}
that was introduced for $\mathcal{N}=4$ sYM in \cite{Ferro:2022abq}. It is defined by
\begin{equation}\label{eq:loop-mom-def}
\ell_l=\frac{\sum\limits_{a<b}(ab)_{D_l} \ell_{ab}^\star}{\sum\limits_{a<b}(ab)_{D_l}\langle ab\rangle}\,,
\end{equation}
where the matrix $D_l$ is an element of Grassmannian space $G(2,2k)$ and $(ab)_{D_l}$ are $2\times 2$ minors of the matrix $D_l$. Moreover, we define
\begin{equation}
\ell_{ab}^\star=	\sum_{c=b+1}^n(-1)^c\lambda_a\langle bc\rangle \lambda_c-\sum_{c=a+1}^n(-1)^c\lambda_b\langle ac\rangle \lambda_c\,.
\end{equation}
The image $\ell_l=\Phi_\lambda(D_l)$ is generically not a symmetric matrix, and therefore in order to adapt this construction to the ABJM theory, we need to restrict to a subsets of matrices $D_l$ which result in three dimensional off-shell momenta. This imposes additional constraints on $D_l$, which correspond to the symplectic condition described in \cite{He:2023rou}.

We define the loop momentum amplituhedron $\mathcal{A}_{2k}^{(L)}$ as the image of a particular subset $\mathcal{D}_L\subset G(2,2k)\times\ldots\times G(2,2k)$ obtained as follows. We take a matrix $C\in OG_+(k)$ such that $\lambda=\phi_\Lambda(C)$ and construct its T-dual version $\check{C}$ (see \cite{Arkani-Hamed:2009kmp}). Then we say that $(D_1,\ldots,D_L)\in \mathcal{D}_L$ if the matrices $(\check{C},D_{l_1},\ldots,D_{l_p})$ are positive for all subsets $\{l_1,\ldots,l_p\}\subset \{1,\ldots,L \}$. Finally, we can define
\begin{equation}
\mathcal{A}_{2k}^{(L)}=\Phi_\lambda(\mathcal{D}_L)\,.
\end{equation}

\subsection{One loop}

We start our exploration of the loop momentum amplituhedron by examining the one-loop geometry $\smash{\mathcal{A}_{2k}^{(1)}}$, which is relatively basic. One straightforward fact that can be derived from definition \eqref{eq:phi_lambda} is that $\smash{\mathcal{A}_{2k}^{(1)}\subset \mathcal{K}^{\leq0}_{x_a}}$, which means that all point $\smash{\ell\equiv x\in \mathcal{A}_{2k}^{(1)}}$ are space-like separated from all points $x_a$. What is far less obvious is the fact that all points of the loop momentum amplituhedron sit in the compact part $\Delta_{2k}(x_a)$ of $\mathcal{K}^{\leq0}_{x_a}$. We have performed extensive numerical tests to confirm that they are actually equal:
\begin{equation}
\mathcal{A}_{2k}^{(1)}=\Delta_{2k}^{(m)}\,,
\end{equation}
where the geometry on the right hand side depends on the choice of tree-level $\lambda$. Most importantly, for all points $\lambda$ in the same chamber, the loop geometry $\mathcal{A}_{2k}^{(1)}$ looks the same, confirming that the geometry factorises as in \eqref{eq:factorisation_chambers}.  

Knowing the geometry, we can find its canonical differential form. From the factorisation property, we immediately get that:
\begin{equation}
A_n^{(1)}=\sum_m \Omega_{n,m}^{(0)}\wedge \Omega_{n,m}^{(1)}\,,
\end{equation}
where $\Omega_{n,m}^{(0)}$ is the tree-level form associated to the chamber $C_m$.
In the following we will derive explicit expressions for $\Omega_{n,m}^{(1)}=\Omega(\Delta_{2k}^{(m)})$. 
 Since $\Delta_{2k}^{(m)}$ is just a curvy version of a simple polytope in three-dimensions, the canonical form should naively be just the sum over vertices of the dlog forms for all facets that meet at that vertex
\begin{equation}\label{eq:formabc}
\Omega_{naive}(\Delta_{2k}^{(m)})=\sum_{(abc)\in \mathcal{V}(\Delta_{2k}^{(m)})}\sigma_{abc}\,\omega_{abc}\,,
\end{equation}
where
\begin{align*}
    \omega_{abc}&=\mathrm{d}\log(x-x_{a})^2\wedge \mathrm{d}\log(x-x_{b})^2\wedge \mathrm{d}\log(x-x_{c})^2\,.
\end{align*}
The signs $\sigma_{abc}$ can be found by demanding that the form is projective, see \cite{Arkani-Hamed:2017mur}. We emphasize that the differential forms $\omega_{abc}$ separately are not dual conformally invariant, however, the final answer \eqref{eq:formabc} is. The difference in our case compared to the story of simple polytopes is that the differential form \eqref{eq:formabc} also has support outside of $\Delta_{2k}^{(m)}$. More precisely, it has a non-zero residue at all points $p_{abc}^{\pm}$, while only one of them is a vertex of $\Delta_{2k}^{(m)}$. Since
\begin{equation}
\mathop{\mathrm{Res}}\limits_{x=p_{abc}^\pm}\omega_{abc}=1\,,
\end{equation}
we need to find a form that contributes with opposite signs on the two points. The natural candidate is the triangle integrand
\begin{align}
&\omega^{\triangle}_{abc}=\frac{4\sqrt{x_{ab}^2x_{bc}^2x_{ac}^2}\mathrm{d}^3x}{(x-x_a)^2(x-x_b)^2(x-x_c)^2}\\
&=\pm\mathrm{d}\log\frac{(x-x_a)^2}{(x-p_{abc}^\pm)^2}\wedge\mathrm{d}\log\frac{(x-x_b)^2}{(x-p_{abc}^\pm)^2}\wedge\mathrm{d}\log\frac{(x-x_c)^2}{(x-p_{abc}^\pm)^2}\nonumber
\end{align}
for which
\begin{equation}
\mathop{\mathrm{Res}}\limits_{x=p_{abc}^\pm}\omega^{\triangle}_{abc}=\pm1\,.
\end{equation}
Therefore, the form that is supported only on $\Delta_{2k}^{(m)}$ is
\begin{equation*}
\Omega(\Delta_{2k}^{(m)})=\sum_{(abc)\in \mathcal{V}(\Delta_{2k}^{(m)})}\sigma_{abc}\,\omega^\pm_{abc}=\sum_{(abc)}\sigma_{abc}\,(\omega_{abc}\pm\omega_{abc}^{\triangle})\,,
\end{equation*}
where the relative sign in the bracket depends on which solution of the triple intersection is a vertex of the geometry. Performing case-by-case studies, one finds that all intersections with even labels have negative sign, while all with odd labels positive sign. Therefore, we conjecture that the one-loop ABJM integrand for the positive branch for any $n=2k$ is \eqref{eq:oneloopall}. The formula agrees with the one provided in \cite{He:2023rou} for $n=4,6,8,10$ \footnote{It would also be interesting to compare our result to the all-multiplicity one-loop formula found in \cite{Bianchi:2012cq} for a subset of component amplitudes.}.

As argued in \cite{He:2023rou}, the complete $2k$-point ABJM integrand is given by a sum over $2^{k-2}$ different branches. While our current construction of the geometry gives the integrand only for the positive branch, the integrands for other branches can be obtained from \eqref{eq:oneloopall} by making use of the parity operations introduced in \cite{He:2023rou}. These effectively interchange certain $p_{abc}^+\leftrightarrow p_{abc}^-$, while keeping all points $x_a$ unchanged. Therefore, we can think of other branches as having exactly the same shape $\Delta_{2k}^{(m)}$ as the positive branch, but with some of the triple intersection points swapped. It is clear that such a parity operation will only flip the signs of the corresponding triangle integrands, while leaving the rest of the form unchanged. Therefore, if one is able to classify all parity operations for given $n$, the full amplitude can be derived from our geometric construction by summing over these relabelled geometries. 

\subsection{More loops}
In this section we will have a first look at implications of our construction for the $L$-loop problem for $L>1$. In this case, each off-shell loop momentum $\ell_l$ is space-like separated from all points $x_a$ and therefore sits inside $\Delta_{2k}^{(m)}$. However, we also need to impose mutual positivity constraints that translate into the requirement that every pair of loop momenta is space-like separated $(\ell_{l_1}-\ell_{l_2})^2<0$ for all ${l_1},{l_2}=1,\ldots,L$. 

\begin{figure}[t!]
	\centering
	\begin{subfigure}[t]{0.2\textwidth}
		\centering
		\includegraphics[scale=0.3]{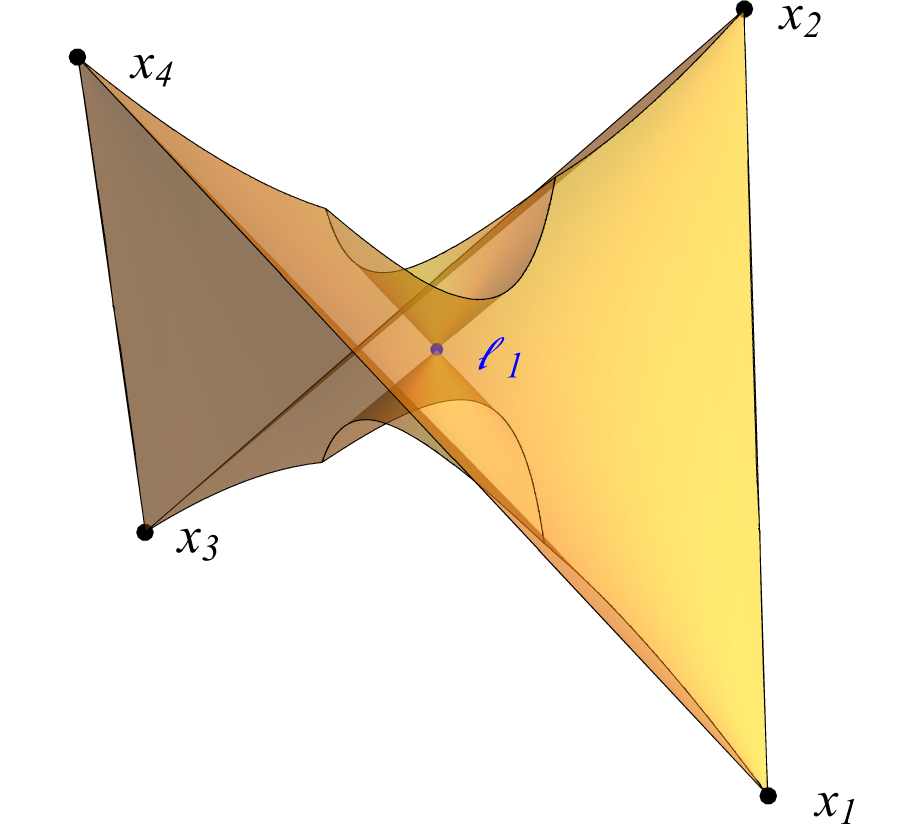}
		\caption{\label{fig:twoloop}}
	\end{subfigure}
	\hfill
	\begin{subfigure}[t]{0.2\textwidth}
		\centering
		\includegraphics[scale=0.4]{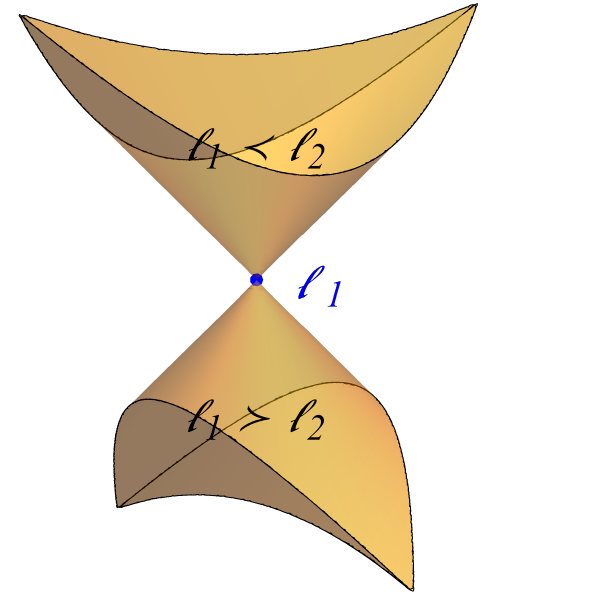}
		\caption{\label{fig:halflightcones}}
	\end{subfigure}
	\caption{The positive (a) and negative (b) part of the geometry for $n=4$ for fixed $\ell_1$.}
\end{figure}

This is particularly simple for $n=4$ at two loops, where for a fixed position of momentum $\ell_1$, the geometry accessible to the momentum $\ell_2$, depicted in figure \Cref{fig:twoloop}, does not depend on the position of $\ell_1$.  
We are therefore interested in the region inside $\Delta_4$ which sits outside the lightcone of $\ell_1$. Importantly, the latter intersects the lightcones of points $x_1$ and $x_3$ in the future and the lightcones of points $x_2$ and $x_4$ in the past. It is not obvious how to directly find the canonical differential form of this region. We will however circumvent this problem by considering negative geometries, see \cite{Arkani-Hamed:2021iya,He:2022cup}. It is clear that 
\begin{equation}
\mathcal{A}_4^{(2)}\cup \mathcal{R}_4=\mathcal{A}_4^{(1)}\times \mathcal{A}_4^{(1)}\,,
\end{equation}
where $\mathcal{R}_4=\{(\ell_1,\ell_2)\in \Delta_4\times \Delta_4: (\ell_1-\ell_2)^2>0)\}$. This region further decomposes into $\mathcal{R}_4=\mathcal{R}_{4,\ell_1\prec \ell_2}\cup \mathcal{R}_{4,\ell_1\succ \ell_2}$, where $\mathcal{R}_{4,\ell_1\prec \ell_2}$ (resp $\mathcal{R}_{4,\ell_1\prec \ell_2}$) contains all points for which $\ell_1$ is in the past (resp. future) of $\ell_2$. The boundary structure of these two regions is significantly simpler then the one of $\mathcal{A}_4^{(2)}$, see figure \ref{fig:halflightcones}.
In particular, the only boundaries accessible by momentum $\ell_1$ are $(\ell_1-\ell_2)^2=0$, $(\ell_1-x_2)^2=0$ and $(\ell_1-x_4)^2=0$ (resp. $(\ell_1-\ell_2)^2=0$, $(\ell_1-x_1)^2=0$ and $(\ell_1-x_3)^2=0$). By comparing with known results, there is a natural differential form that we can associate to each of these regions:
\begin{equation*}
\Omega(\mathcal{R}_{4,\ell_1\prec \ell_2})=\frac{x_{13}^2x_{24}^2\,\mathrm{d}^3\ell_1\wedge\mathrm{d}^3\ell_2}{(\ell_1-x_2)^2(\ell_1-x_4)^2(\ell_1-\ell_2)^2(\ell_2-x_1)^2(\ell_2-x_3)^2}
\end{equation*}
and $\Omega(\mathcal{R}_{4,\ell_1\succ \ell_2})=\Omega(\mathcal{R}_{4,\ell_1\prec \ell_2})|_{\ell_1\leftrightarrow \ell_2}$.
Therefore, the two-loop integrand is given by
\begin{equation}
\Omega_4^{(2)}=\Omega^{(1)}_1(\ell_1)\wedge\Omega^{(1)}_1(\ell_2)-\Omega(\mathcal{R}_{4,\ell_1\prec \ell_2})-\Omega(\mathcal{R}_{4,\ell_1\succ \ell_2})\,.
\end{equation}
This agrees with \cite{Chen:2011vv}.

\begin{figure}
\begin{center}
	\includegraphics[width=41mm]{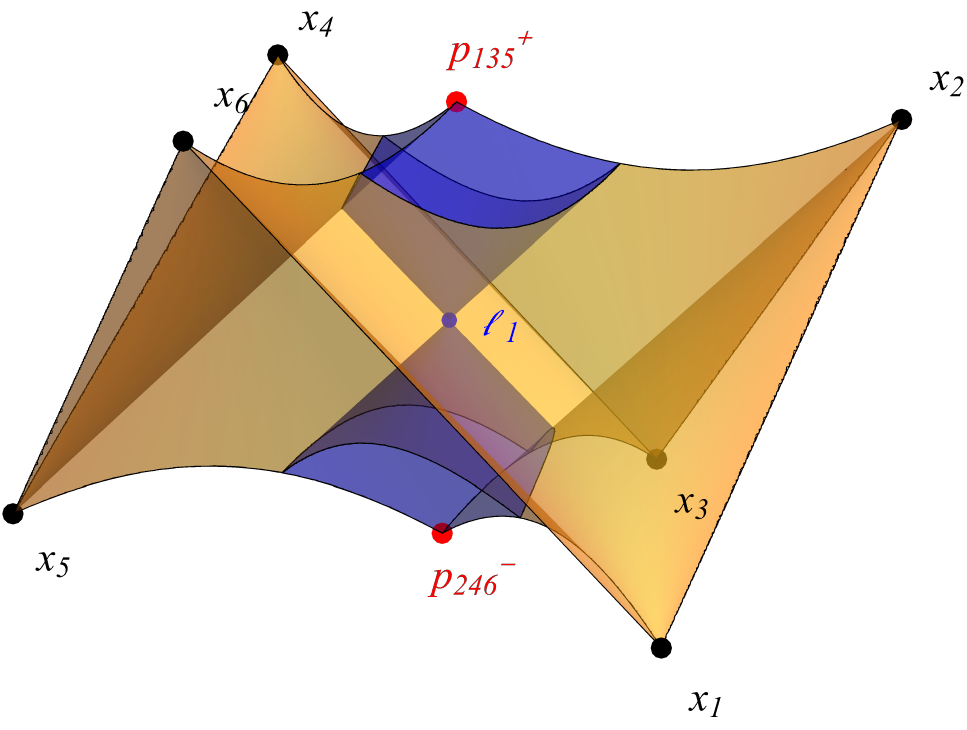}%
	\includegraphics[width=44mm]{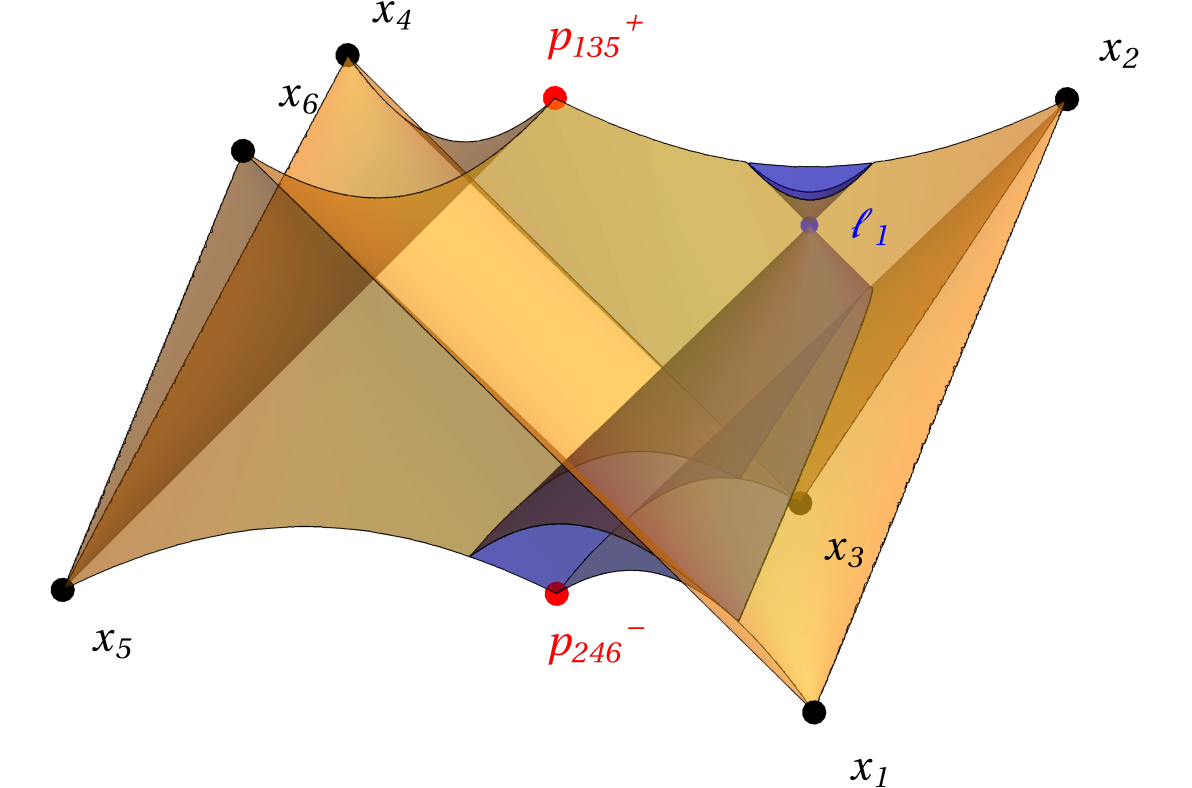}
\end{center}
\caption{Two of the 13 possible regions in the two-loop geometry  for fixed position of $\ell_1$ for $n=6$.}
\label{fig:regionsn6}
\end{figure}

For higher number of points at two loops, we observe that, when going to negative geometries, we can again define two regions: $\mathcal{R}_{n,\ell_1\prec \ell_2}$ and $\mathcal{R}_{n,\ell_1\succ \ell_2}$, where loop momenta are time-like separated and time-ordered. Unlike for $n=4$, we get different regions for $\ell_2$ depending on the position of $\ell_1$. However, there is a simple classification of all these regions. We focus first on $\mathcal{R}_{6,\ell_1\prec \ell_2}$ and notice that for fixed $\ell_1$ the only boundaries for momentum $\ell_2$ are at $(\ell_1-\ell_2)^2=0$ and at the lightcones of points $x_a$ that intersect the lightcone of $\ell_1$ in the future. There are four possibilities: 
\begin{align}
&\{\mathcal{I}_{\ell_1}\cap\mathcal{I}_{x_1}\neq\emptyset,\mathcal{I}_{\ell_1}\cap\mathcal{I}_{x_3}\neq\emptyset\}\,,\label{eq:n6-x1x3}\\
&\{\mathcal{I}_{\ell_1}\cap\mathcal{I}_{x_1}\neq\emptyset,\mathcal{I}_{\ell_1}\cap\mathcal{I}_{x_5}\neq\emptyset\}\,,\\
&\{\mathcal{I}_{\ell_1}\cap\mathcal{I}_{x_1}\neq\emptyset,\mathcal{I}_{\ell_1}\cap\mathcal{I}_{x_5}\neq\emptyset\}\,,\\
&\{\mathcal{I}_{\ell_1}\cap\mathcal{I}_{x_1}\neq\emptyset,\mathcal{I}_{\ell_1}\cap\mathcal{I}_{x_3}\neq\emptyset,\mathcal{I}_{\ell_1}\cap\mathcal{I}_{x_5}\neq\emptyset\}\,,\label{eq:n6-x1x3x5}
\end{align}
as depicted in figure \ref{fig:regionsn6}.
Similar analysis holds true for boundaries for $\ell_1$ when $\ell_2$ is fixed: the four possibilities are $\{2,4\}$, $\{2,6\}$, $\{4,6\}$ or $\{2,4,6\}$. 
We found that there are 13 allowed regions in this geometry (notice that region (\{2,4\}, \{1,5\}) and its two cyclic rotations are not allowed), which correspond to the bipartite graphs in \cite{He:2023rou}. Therefore, it should be possible to rewrite the answer found there such that each term matches one of these 13 regions. Similar structure is also present in the $n=8$ two-loop answer, and it naturally follows from our construction, since it reflects which lightcones intersects $\mathcal{I}_{\ell_1}$ and $\mathcal{I}_{\ell_2}$ in the past/future. At the moment we do not have a good understanding of how to associate differential forms to this regions and leave it for future work.

Finally, by moving to higher loop orders, one  can use negative geometries and study configurations of loop momenta inside the region $\Delta_{n}^{(m)}$ that are (partially) time-ordered in three-dimensional Minkowski space, as suggested in \cite{He:2022cup}. Our construction provides a simple geometric picture that can be used to organise the calculations based on the causal structure of the corresponding configuration of loop momenta.

\section{Conclusions and outlook}
In this paper we defined the spinor helicity version of the ABJM amplituhedron and, by translating it to the space of dual momenta, we found a surprisingly simple geometry associated with the causal structure of configurations of point in three-dimensional Minkowski space. This allowed us to find a formula for all integrands at one loop, and shed some light on the structure of the answer at two loops and beyond. 

Intriguingly, if we knew nothing about amplituhedra, we could still define $\Delta_n(x_a)$ as a compact region in the Minkowski space that is space-like separated from a null polygon with $n$ vertices. By studying this region we would rediscover the structure of scattering amplitudes in ABJM theory! However, it is far from obvious why ABJM theory is selected from all possible three-dimensional quantum field theories. Finding the answer to this question might shed light on generalisations of our construction beyond ABJM. 

There are many interesting avenues to follow based on this new picture of scattering in ABJM. The most urgent one is to better understand the geometry itself and in particular to provide a constructive way to derive its differential forms. It will be crucial to check whether the structure of the answer that we saw at one loop can be systematically generalised to higher loops, promising all multiplicity answers for ABJM loop integrands.

Finally, the ABJM momentum amplituhedron is a reduction of the  momentum amplituhedron in $\mathcal{N}=4$ sYM. A natural question that arises is whether a similar basic geometry lives in the four-dimensional space of dual momenta in $(2,2)$ signature.

\begin{acknowledgments}
The authors would like thank Song He, Yu-tin Huang and Chia-Kai Kuo for useful discussions on their work. We would like to thank the Dublin Institute for Advances Studies for hosting the workshop ``Amplituhedron at 10'', which provided the initial motivation for this work. 
\end{acknowledgments}

\bibliography{ABJM_mom_ampl}
\end{document}